# Atomistic simulations of dislocation mobility in Al, Ni and Al/Mg alloys

David L. Olmsted[*], Louis G. Hector, Jr.[†], W. A. Curtin[*] and R. J. Clifton[*]

[*]Division of Engineering, Brown University
Providence, RI 02912

[†]General Motors Technical Center
30500 Mound Rd., Warren, MI 48090

## ABSTRACT

Dislocation velocities and mobilities are studied by Molecular Dynamics simulations for edge and screw dislocations in pure aluminum and nickel, and edge dislocations in Al-2.5%Mg and Al-5.0%Mg random substitutional alloys using EAM potentials. In the pure materials, the velocities of all dislocations are close to linear with the ratio of (applied stress)/(temperature) at low velocities, consistent with phonon drag models and quantitative agreement with experiment is obtained for the mobility in Al. At higher velocities, different behavior is observed. The edge dislocation velocity remains dependent solely on (applied stress)/(temperature) up to approximately 1.0 MPa/K, and approaches a plateau velocity that is lower than the smallest "forbidden" speed predicted by continuum models. In contrast, above a velocity around half of the smallest continuum wave speed, the screw dislocation damping has a contribution dependent solely on stress with a functional form close to that predicted by a radiation damping model of Eshelby. At the highest applied stresses, there are several regimes of nearly constant (transonic or supersonic) velocity separated by velocity gaps in the vicinity of forbidden velocities; various modes of dislocation disintegration and destabilization were also encountered in this regime. In the alloy systems, there is a temperature- and concentration-dependent pinning regime where the velocity drops sharply below the pure metal velocity. Above the pinning regime but at moderate stresses, the velocity is again linear in (applied stress)/(temperature) but with a lower mobility than in the pure metal.

# 1. INTRODUCTION

Since its existence was independently deduced in 1934 by Taylor, Orowan, and Polyani[1], the dislocation has been the subject of intense theoretical and experimental study. Of particular interest is dislocation motion in metals, which governs many of the macro-scale properties of metals during plastic deformation. Plasticity models have been formulated at a variety of length scales that have varying connections to the physical nature of dislocations. For example, continuum crystal plasticity makes use of the established dislocation slip systems in a single crystal structure to formulate constitutive rules (i.e. stress vs. strain rate) at the macroscale[2]. More recently, strain gradient plasticity models have been developed to account for size-dependent effects in plasticity that are absent from standard continuum models[3]. At the same time, discrete dislocation models have been developed wherein the motion of many interacting dislocations is followed explicitly in order to predict, for example, single-crystal stress-strain curves as a function of strain rate[4,5,6,7,8,9,10]. In such discrete dislocation models, the dislocations are continuum line defects whose dynamics are controlled by a driving force, the Peach-Koehler force, $F$ [11], and an assumed dislocation mobility law. The motion of these dislocations may be subject to various constraints, including pinning sites such as dislocation junctions and precipitates. The dislocation mobility is typically assumed to be described by

$$Bv = F \qquad (1)$$

where v is the dislocation velocity and $B$ is the drag or damping coefficient ($B^{-1}$ is the mobility factor)[6,8,9,10]. For more fully dynamic simulations that include dislocation inertia, the same form of damping has been included as a retarding force[7]. The physics described by Eq. (1) is a resistance to dislocation motion in the form of a viscous friction force; theoretical studies indicate that such forces arise from dislocation interactions with lattice phonons and with electrons[12]. As more quantitative results are sought from crystal plasticity and discrete dislocation models, better understanding of the origin, accuracy, and range of validity of the underlying mobility law, including issues such as the actual magnitude and dependence of the damping coefficient $B$ on temperature and dislocation character, will be required. Beyond issues related to "ordinary" stresses inherent in metal processing or in-service deformation in various structural applications, dislocation motion under ultra-high stresses occurring under intentional and unintentional shock conditions and in high rate experiments is expected to be outside the range of validity of the linear mobility law, thus raising further questions about dislocation motion under such extreme loadings. Finally, in many metallic systems such as 5XXX Al-Mg alloys, substantial strengthening can be provided by substitutional solute atoms. The strengthening and rate dependence of solute hardening is of fundamental importance to metal alloy design and performance, but a clear understanding of dislocation motion in pure elemental substances such as Al is desirable as a basis for the study of more complex systems.

Considerable prior work on modeling and simulation of dislocation dynamics has been carried out in the last fifty years. Key aspects of those efforts are discussed in the next section. In the present work, we build on the prior work and bring the power of large-scale simulation methods to bear on the study of dislocation dynamics. Specifically, we report results from atomistic simulation studies of edge and screw dislocation motion in aluminum (fairly isotropic) and nickel (more anisotropic) over temperatures ranging from 100K to 500K and stresses ranging from 5 MPa to 4 GPa (see Table 1 for relevant material properties). Since the processes by which dislocation motion is damped are different at different velocities and in different material environments, we also investigated the motion of a single edge dislocation in Al-2.5%Mg and Al-5.0%Mg wherein Mg is randomly substituted. For this purpose, we chose temperatures ranging from 300K to 700K, and stresses ranging from 5 MPa to 210 MPa. These choices allowed us to explore the influence of two drag mechanisms: solute pinning/de-pinning energetics and lattice phonon damping over a wide range of velocities. Our results confirm various scalings and observations made in earlier work, and uncover some new phenomena as discussed below. In particular, the mobility for edge dislocations scales with resolved shear stress divided by the temperature ($\sigma/T$) over a wide range beyond the expected linear regime governed by phonon-damping while the mobility for screw dislocations scales rather differently and in a manner similar to that expected from radiation-damping models

The remainder of this report is organized as follows. In Section 2, we briefly review the relevant theoretical background for the dislocation dynamics phenomena studied herein and the prior work on atomistic simulations. In Section 3, we describe the model geometry and simulation details associated with our molecular dynamics study. In Section 4, we present our results and discuss them within the framework of existing models and prior observations. Section 5 contains a summary of the most significant results of this work.

## 2. Theoretical Models and Prior Simulation Studies

Three areas of existing theoretical work are relevant to our simulations: phonon damping, continuum elasticity predictions of the velocities at which the solution for a dislocation in uniform motion becomes singular (i.e. the so-called "forbidden velocities"), and radiative damping (at velocities below the lowest forbidden velocity) resulting from the discrete crystal lattice.

Existing thermal phonons in a crystal lattice scatter from a moving dislocation, thereby damping the dislocation motion. For velocities small compared to elastic wave speeds, phonon damping models predict a steady state dislocation velocity v of the form

$$B(T)v = b\sigma \qquad (2)$$

where b is the magnitude of the Burgers vector, σ is the resolved shear stress, and B(T) is a damping coefficient. The details of the temperature dependence of B depend on the specific model, but for temperatures high compared to the Debye temperature $T_D$, B is linear in T [13][14]. Phonon damping due to existing thermal phonons is proportional to the phonon density which, in turn, is proportional to T for $T > T_D$. Because no quantum effects appear in the simulations, all temperatures are high compared to $T_D$ for the classical potential. In the original Leibfried model, the phonon damping contribution to B is [13][15]

$$B = \frac{s\bar{e}}{10c_t} \qquad (3)$$

where $\bar{e}$ is the phonon energy density, $c_t$ is the transverse wave speed, and $s$ is the dislocation scattering width, argued to be of the order of the Burgers vector. For our classical simulations, $\bar{e} = 3Nk_B T/V$, with N the number of atoms and V the total volume, because the thermostat keeps the average kinetic energy equal to half this value and all of the thermal energy is phonon energy.

For steadily moving dislocations, continuum linear elasticity theory provides solutions similar to those for stationary dislocations[11]. The dislocation energies predicted from these solutions (ignoring the long range logarithmic divergence) diverge at certain velocities, which for isotropic elasticity and for high symmetry directions of motion in anisotropic elasticity are the wave speeds in the material[16][17][18][11]. Therefore, a dislocation cannot travel steadily at these so-called "forbidden velocities"[19]. However, as first discovered by Eshelby[20][11][21], there may be solutions for velocities that exceed the lowest forbidden velocity, and for isotropic elasticity he found a steady state non-radiative solution at a velocity equal to $\sqrt{2}$ times the shear wave speed. Stroh[18] noted, however, that the behaviors of the screw and edge dislocation at (and approaching) the transverse wave speed are not identical: the energy diverges as a -3/2 power for the edge dislocation and as a -1/2 power for the screw dislocation. In anisotropic linear elasticity, three forbidden velocities exist which, in high symmetry situations, correspond to three wave velocities for propagation in the direction of dislocation motion. Stroh[18] further implies that the anisotropic screw dislocation behaves like the isotropic screw dislocation, in that resonance does not occur at the forbidden velocity as long as the Burgers vector possesses certain symmetry properties (satisfied by the fcc materials considered here). It is therefore expected that the behavior of edge and screw dislocations will be different as they approach the forbidden velocity.

Continuum elasticity does not take into account the discrete nature of the underlying crystal. Low frequency waves propagate at wave speeds calculated by continuum elasticity, but higher frequency waves travel at lower speeds. Therefore, a moving dislocation can be expected to radiate high frequency phonons at velocities below the forbidden velocity, as was first analyzed by Eshelby[20] for the case of a screw dislocation in an isotropic medium using an isotropic dispersion law and a Peierls-Nabarro dislocation. The dislocation begins to radiate when its velocity exceeds the phase velocity, $v_m$, of the slowest phonon mode in the first Brillouin zone. For a velocity v slightly greater than $v_m$, the radiative damping is proportional to $(v - v_m)^{3/2}$. Later work[22][23] suggests that because of Umklapp processes (i.e. phonon scattering that changes the total crystal momentum by a reciprocal lattice vector) there is no slowest mode at which a dislocation can radiate, and that radiative damping should be important even at low velocities (see also Refs. 24 and 25). An additional mechanism of radiation is associated with the motion of a dislocation through the Peierls energy landscape. The dislocation should slow down as it climbs the landscape hills and subsequently speed up as it comes back down the landscape hills, giving a small oscillatory motion superimposed on the overall motion. The core structure can also be expected to change slightly between the tops of the hills and the valleys. Both of these oscillations emit some radiation and hence cause damping, but both are expected to require only forces small compared to the Peierls force to overcome them[26]. Because the Peierls barrier is low in pure fcc metals, this is not expected to be a substantial effect in this case, and will not be considered further.

Basic features of dislocation motion can also be uncovered by atomistic simulations. Among the many features of dislocations that have been studied with atomistic simulations are the structure of dislocation cores[27], barriers to dislocation motion[28], and the structure of dislocation junctions[29]. Much less work has been reported on the actual atomistic dynamics of dislocation motion, and we discuss much of the three-dimensional work in the next few paragraphs.

Daw et al.[30] reported on molecular dynamics (MD) simulations of edge dislocations in fcc nickel with and without hydrogen interstitial impurities. For pure nickel, they observed, at a single stress at room temperature, a steady velocity in good agreement with Leibfried's phonon damping model[13]. A saturation velocity at high stress was observed at about 60% of the lowest forbidden velocity of $2.33 \times 10^3$ m/s for their Ni potential (see Table 1).

Gotelueschen[31] investigated the motion of edge and screw dislocations in an fcc Lennard-Jones crystal[32], and observed subsonic and transonic velocities for the edge dislocation and subsonic velocities for the screw dislocation. A regime at lower stress was noted where the velocity of edge dislocations was approximately linear with stress. Damping coefficients were found to vary linearly with temperature, implying a velocity proportional to applied stress divided by temperature. For the screw dislocation, the low stress data was not linear; however, damping coefficients fit to the data were well fit as linear with temperature, and the resulting damping coefficients were roughly 2.5 times those for the edge dislocation. At higher stresses, a simulation using a ramped stress showed a leveling off of the velocity at about 90% of the lowest forbidden velocity, and then a jump to a transonic velocity somewhat below the second forbidden velocity (the higher of the two transverse wave speeds). Ramping down the stress produced a hysteresis loop, with the velocity eventually dropping back into the subsonic regime. For the screw dislocation, high velocity simulations were not attempted. Our work below will report very similar results for more-realistic systems.

Gumbsch and Gao[33] performed micro-canonical MD simulations of edge dislocation motion in elastically isotropic bcc tungsten using a Finnis-Sinclair potential[34], and observed subsonic, transonic, and supersonic velocities. They observed a maximum subsonic velocity of about $0.7c_T$, where $c_T$ is the transverse wave speed, and a preferred transonic velocity around $\sqrt{2}c_T$ ($1.38c_T$ to $1.5c_T$). Even at $\sqrt{2}c_T$, where the linear elasticity solution indicates no radiation[20], the moving dislocations did radiate, which was attributed to nonlinear core effects. Rosakis[35] used an augmented Peierls-Nabarro model to study the possibility of transonic and supersonic dislocation motion that provided qualitative agreement with the Gumbsch and Gao[33] data but without predicting any preferred velocity below the transverse wave speed. Rosakis' results suggest that the maximum subsonic velocity found by Gumbsch and Gao is the velocity at the stress above which transonic motion is stable.

Bhate et al.[36] investigated motion of edge dislocations in fcc aluminum using the Ercolessi and Adams EAM potential[37] employed here and observed subsonic and transonic velocities. They noted four regimes of behavior as stress increases. At the lowest stress and highest temperature the dislocation does not move, which was interpreted as being controlled by the Peierls barrier. At higher stresses, the dislocation moves with a velocity that decreases as temperature increases, which was attributed to thermal phonon interactions. At yet higher stresses, the dislocation velocity reaches a plateau at about 2.1 nm/ps, lower than the lowest forbidden velocity and essentially independent of stress and temperature, consistent with results from Grotelueschen[31] and Gumbsch and Gao[33], but not in accordance with Rosakis' model[35]. At the highest stress, a velocity jump to roughly $\sqrt{2}c_T$ was noted, in accordance with Eshelby's prediction.

Each of the above studies of edge dislocation motion suggests a saturation velocity at high stresses that is below the smallest shear wave speed, $c_{min}$. The saturation values vary from about $0.7c_{min}$ for the bcc material to about $0.9c_{min}$ for the fcc Lennard-Jones system. Such a saturation was not observed, but also not ruled out, for the screw dislocation in the Lennard-Jones model;

however, the largest velocities achieved at finite temperature were $0.7c_{min}$-$0.8c_{min}$ at zero temperature[31].

With this background, we now turn to describe simulations of edge and screw dislocation dynamics in Ni and Al and edge dislocation dynamics in Al-Mg alloys over a wide range of temperatures and stresses.

## 3. SIMULATIONS

The MD simulations model the motion of a nominally straight edge or screw dislocation of infinite length moving under a constant applied stress at a constant temperature. With some exceptions, the simulation cell is pre-strained at an elastic strain that is somewhat less than that implied by the stress so that the initial stress wave from the boundaries is reduced. The Ercolessi-Adams EAM potential[37] was used in the aluminum simulations and was augmented by the Liu et al. EAM potentials[38][39] for the aluminum-magnesium simulations. For the nickel simulations, a potential due to Daw et al.[30] was employed. The simulations were run at constant temperature using a Hoover thermostat applied to the entire system.

Here, we briefly describe the simulation cell geometry, construction, and testing since cell size and shape actually have an influence on the dislocation dynamics that must be qualitatively understood and minimized. The configuration of the simulations, which is shown schematically in Fig. 1 as a rectangular simulation cell that contains the dislocation, is similar to that used by Bhate et al.[36]. A single dislocation line (of either edge or screw character) oriented along the z-direction is introduced in a slip plane perpendicular to the y-direction. The simulation cell has periodic boundary conditions in the z-direction, with length $L_z$. The cell also has periodic boundary conditions in the direction of dislocation motion, with length $L_x$. In the y-direction, the simulation cell is of length $L_y$ with non-periodic boundary conditions: the atoms within several layers at the top and the bottom surface have fixed y-coordinates, so that these layers stay perfectly flat. These surfaces thus reflect acoustic waves and cannot sustain all of the surface waves available to a free surface. The imposed boundary conditions give rise to an effective linear array of dislocations spaced by a distance $L_x$ moving in a constrained thin film. The cell dimensions for each of the three systems, along with the total number of atoms in each cell, are listed in Table 2. The system in Fig. 1 was then subject to an applied shear stress as shown.

For the aluminum edge dislocations, a series of initial calculations involving different cell sizes were performed to test for size effects. These calculations were conducted using simulation cells having the smallest possible length $L_z \approx 5$Å, which represents 6 atomic planes. For the sake of comparison, some simulations were conducted using traction free boundary conditions in the normal direction at the two y-surfaces. These simulations were conducted at low temperature (e.g. 10 K) with applied stresses of 60 MPa and 250 MPa; such conditions lead to high subsonic dislocation velocities that might be expected to be more sensitive to boundary conditions than lower velocities. The simulations with zero normal traction boundary conditions were more sensitive to the size of the simulation cell, and took longer to reach a steady velocity in some cases. When extrapolated to large cell size while keeping the aspect ratio constant, however, the velocities for the zero normal traction boundary conditions and for the zero normal displacement boundary conditions appeared to converge to the same limit. For fixed normal displacement boundary conditions at the surface, the velocities for the simulation cell size used for the data presented herein were slightly lower than those for the larger system sizes (at constant aspect ratio), but were within 5% of the values extrapolated to large system size. It should be noted, however, that both boundary conditions tested at the surfaces are reflective. We did not conduct any simulations with so-called "non-reflective" boundary conditions.

## 4. RESULTS AND DISCUSSION

4.1 Dislocation Mobility

We first address the issue of dislocation mobility in pure materials at low velocities. Figure 2 shows the dislocation velocity as a function of the applied resolved shear stress divided by the temperature (hereafter $\sigma/T$) for an edge dislocation in pure nickel (open triangles) and an edge dislocation in pure aluminum (solid triangles). The velocity is very nearly linear in $\sigma/T$ up to about 0.3 MPa/K. In this range both the linearity and the fact that the velocity is a function of $\sigma/T$ alone, rather than of the applied stress and the temperature separately, is clear. This result agrees with the scaling predicted by phonon damping theories[13][14]. Figure 3 shows the dislocation velocity as a function of $\sigma/T$ for screw dislocations in pure Ni (open triangles) and in pure Al (solid triangles). Again, the velocity depends only on $\sigma/T$ up to about 0.4 MPa/K, and coincidentally the Al and Ni data are essentially indistinguishable over this range. The velocity is roughly linear in $\sigma/T$, but not nearly as linear as the edge dislocation velocity. For $\sigma/T$ larger than 0.6 MPa/K, the behavior is no longer linear and, in contrast to the edge dislocation data, it is no longer a function of $\sigma/T$ alone. Thus, at intermediate velocities, the edge and screw exhibit entirely different dependencies on stress and temperature.

Using a least squares procedure, the velocity data in Figures 2 and 3 was fit to the linear form $v = (b/B)\sigma/T$ in the range $\sigma/T < 0.3$ MPa/K to obtain values of $B/T$, which are shown in Table 3. Fits using the data for $\sigma/T < 0.15$ MPa/K are also shown, and are very similar for the edge dislocation results but differ for the screw dislocation due to the non-linear behavior. We find stronger damping, by a factor of 1.6-1.9, for the Al screw dislocation than for the edge, but much less difference between the screw and edge dislocations in Ni. To our knowledge, existing theoretical studies of phonon damping do not offer any prediction of the relative damping for screw and edge dislocations in pure materials.

Quantitatively, we can compare our predictions for $B/T$ and $B$ to various experimental results in Al. The data of Hikata et al.[15] were shown to be fit very well by a model wherein the phonon damping was given by Leibfried's model[13], and by a similar fit using a model wherein phonon damping was given by strain field scattering[15], both within a Debye approximation. To make a model-independent comparison with the data of Hikata et al., we have computed the high-temperature limits of $B/T$ for their fits, yielding $2.2 \times 10^{-8}$ Pa×s/K and $2.3 \times 10^{-8}$ Pa×s/K for the edge and screw, respectively. The $B/T$ measured here (Table 3) for the aluminum edge (screw) dislocation are about twice (three times) this value. Nadgornyi[40] shows two sets of high velocity data for Al, with damping coefficients B at 300K of roughly $2 \times 10^{-5}$ Pa-sec[41] and $6 \times 10^{-5}$ Pa-sec[42], respectively. These compare to a value from Hikata et al. of about $5 \times 10^{-6}$ Pa-sec by extrapolating from their highest data at 250K. Our values for B are about $1.1 \times 10^{-5}$ Pa-sec for the edge and 1.8-2.2 $\times 10^{-5}$ Pa-sec for the screw. Thus, the present MD simulations are within the range of reported experimental values.

We now turn to investigate the mobility in the alloy systems, and limit our investigation to the stress range where linearity in $\sigma/T$ is found in the pure Al. Figure 4 shows the velocity of edge dislocations as a function of $\sigma/T$ for Al with 2.5 at% Mg (open triangles) and 5.0 at% Mg (solid triangles), with the Mg introduced as a random substitutional alloying agent[*]. The dislocation length (i.e. the length of the simulation cell in the direction of the dislocation line) is 20 times the minimal repeat distance, or about 100 Å (Table 2). Each data point represents the average of 8 simulations of different random distributions of Mg atoms, with the velocity in each simulation calculated as the average velocity over the last 60 ps of the total 120 ps simulation time. Above 0.15 MPa/K, the 500K and 700K data show that the velocity is a linear function of $\sigma/T$, so that the damping coefficient for both alloys is linear in temperature as found for pure Al. For the sake of comparison, the linear fit of the pure aluminum edge dislocation velocity with $\sigma/T$ is shown in Figure 4 as the dotted lines. The increase in the damping coefficient with Mg content is 5 to 10 times that implied by the Leibfried model[13][15] based on the changes in lattice constant and elastic moduli, assuming a constant scattering width. Below 0.15 MPa/K, the dislocation velocity is lower than predicted by the linear fit to data above 0.15 MPa/K and also ceases to be a function solely of $\sigma/T$. This "slow" dislocation motion at low stress occurs because the dislocations can be stationary (pinned) for some, or all, of the simulation time. The pinning/de-pinning effect is more pronounced at lower temperatures, as expected. We have examined this "solute strengthening" effect in considerable detail and will report on it in a separate publication. In this pinning/de-pinning regime, the average dislocation velocity will be a function of dislocation length in the simulation, as well as the starting point within the simulation cell, run time, and other factors. At higher stresses where the dislocations are moving freely, the velocity is less dependent on these other factors.

---

[*] The solute atoms are distributed randomly in each plane parallel to the slip plane, with each such plane having the desired concentration of Mg.

At 300K, we observe slightly different behavior. For both Al/5%-Mg and Al/2.5%-Mg, the dislocation velocities are generally lower than those at higher temperature for the same value of $\sigma/T$ and the data exhibits no regime approximating linearity in $\sigma/T$. Thus, the "standard" linear regime caused by phonon damping does not exist. The pinning effect is more severe at the lower temperature; dislocations pinned at local favorable configurations of Mg atoms are thermally released at a much slower rate at the lower temperature, and the regime of zero, or very low, velocity extends to higher stresses at the lower temperature. In addition, however, at stresses above the pinning regime, the 300K data begins rolling over at relatively low velocities or stresses where the 500K and 700K data are still linear; this is particularly evident for the Al/5%-Mg. The retarding mechanisms causing this reduced velocity at high stresses are unknown at present.

4.2 Dislocation Dynamics at High Stresses

We now investigate the dislocation velocity and behavior in the pure materials at stresses above the linear regime. Figures 5 and 6 show the velocity for moving edge dislocations in aluminum and nickel, respectively, as a function of stress only. At stresses above the linear regime, the velocity rolls over toward a plateau velocity that is somewhat below the smallest forbidden velocity. In Figure 2 notice that the dislocation velocity remains a function only of $\sigma/T$ during the roll-over and entrance to the plateau. The temperature and stress independence of the plateau velocity have been reported previously[31][36], but the extent to which the behavior remains purely a function of the ratio $\sigma/T$ has not been noted. Further examination of Figures 5 and 6 reveals that the "plateau" velocity is about 80% of the lowest forbidden velocity for Al and 87% for Ni. However, an actual plateau velocity where the slope of the velocity vs. stress is zero over a measurable range of stress is not clearly observed although the slope becomes extremely small. Furthermore, our data show that at stresses in the neighborhood of 500 MPa for Al and 1500 MPa for Ni, the dislocation velocity gradually increases above the plateau velocity, suggesting that the plateau velocity is a "preferred" velocity for dislocation motion, rather than an asymptotic approach to some forbidden velocity.

A convincing physical interpretation of the plateau velocity has not yet been offered. However, Bhate et al.[36] have suggested that it might be explained by analogy to the lattice dynamics model of Celli and Flytzanis[23] in which there is one or more special velocities where phonon modes exist with equal phase and group velocities in the direction of dislocation motion. The relevant phase velocity determined by Bhate et al. from the phonon dispersion relations of Al was about 25% below the plateau velocity in their simulations of about 2.1 nm/ps, however, and significantly lower than the plateau velocity of 2.6 nm/ps found here, using what we believe to be more appropriate boundary conditions for the problem. We therefore believe that the explanation for the plateau velocity offered by Bhate et al. is not well supported.

As dislocation velocities exceed the plateau value, they are no longer varying as a function of $\sigma/T$ alone (Figures 5 and 6). This is not surprising since the radiation of phonons is expected to become the main damping mechanism as forbidden velocities are approached. Our data suggests that the dislocation velocity is more nearly a function of applied shear stress alone at values above the plateau. In particular, the preferred plateau velocities do not exhibit much temperature dependence. More detailed analysis suggests that the first transonic "preferred" velocity for aluminum is about 5 nm/ps and for nickel it is about 3 nm/ps. The data points for velocities above the plateau, but below these smallest transonic "preferred" velocities, show the average velocity (averaged over a rolling 20 ps period) wandering significantly more, compared with the fairly steady average velocity for the preferred velocities. An example of this behavior is shown in Fig. 7 for Ni at 300K. This behavior is qualitatively consistent with the existence of an actual "forbidden" velocity at (or near) the forbidden velocity predicted by continuum linear elasticity.

At the highest velocities considered, many of the aluminum edge dislocations undergo some type of disintegration. One type of disintegration observed was the complete separation of the two partials. Once the partials separate significantly, they may start to interact more strongly with the nearer partial of the image dislocations (from the periodic boundary conditions). Because the separation between the images in the direction of motion is only about 100 Å, the simulation results cannot be trusted after partial separation. Some simulations for Al showed a substantial period of normal dislocation motion before disintegration, however, and a representative velocity before disintegration is plotted as an open triangle for 324K at 707 MPa and 841 MPa in Figure 5. In some simulations, disintegration occurred at a stress as low as 707 MPa. The Ni edge dislocation simulations, on the other hand, did not show any disintegration of the dislocation for stresses less than 3500 MPa.

Figures 8 and 9 show the dislocation velocity as a function of stress for the Al and Ni screw dislocations, respectively. After the approximately linear regime up to velocities of about 1.4 nm/ps, the slopes decrease substantially. There are two striking differences compared with the edge dislocation results shown in Figures 5 and 6. First, there is no shoulder in the velocity that depends only on $\sigma/T$. Second, there is no subsonic plateau in the velocity at higher stress. However, similar to the edge dislocations, at high stresses the velocity depends primarily on stress and is insensitive to temperature.

Our simulations of the screw dislocation in Al succumb to various forms of damage before any transonic velocities are reached; none of the simulations drove the Al screw dislocations above about 0.87 of the lowest forbidden velocity. In contrast, the Ni screw dislocation could attain transonic velocities without any evidence of a plateau. As far as we are aware, these are the first MD simulations of fcc screw dislocations showing transonic motion. While only a few of the Ni screw dislocation velocities are transonic, with the largest being 106% of the smallest "forbidden" velocity, the jump across the forbidden velocity is hinted at in Figure 9. The "jump" was further investigated by looking at moving averages of the velocity over time for each simulation. Simulations with an average velocity between ~2.3 nm/ps and ~2.8 nm/ps showed greater fluctuations in the velocity over time than did simulations where the average velocity was between 1.6 nm/ps and 2.3 nm/ps or was higher than 2.8 nm/ps. This behavior is consistent with the existence of a velocity, or range of velocities, between 2.3 nm/ps and 2.8 nm/ps, that is especially unfavorable for screw dislocation motion. The lack of a plateau in the velocity is very different behavior from that of the edge dislocations. Perhaps we should expect different behavior for the screw and edge in this regime since the singularity at the forbidden velocity, as predicted by isotropic linear elasticity, is of a different nature for the two dislocation characters[18]. It would thus be interesting to study dislocations with mixed character.

Returning to the subsonic data for the screw dislocations, we find that the screw dislocation data for both Al and Ni can be well-fit using the phenomenological form

$$\sigma = \begin{cases} ATv, & v < v_o \\ ATv + D(v - v_o)^{3/2}, & v > v_o \end{cases} \quad (4)$$

where $A = B/T$ from the linear regime, and $D$ and $v_o$ are fitting parameters, for velocities up to about 80% of the lowest "forbidden" velocity, as shown in Figures 8 and 9. The second expression in Eq. (4) is intended to represent two additive damping terms: the expected phonon damping, proportional to temperature, and a temperature-independent radiative damping with the functional form suggested by Eshelby[20]. Because radiative damping does not involve existing thermal phonons, it should not be strongly temperature dependent, and should depend primarily on dislocation velocity. The fitted $v_o$ values are 1.47 nm/ps for Al and 1.52 nm/ps for Ni, neither of which corresponds to the velocities of any relevant (i.e. slowest) phonon modes in the full wave spectrum of the bulk materials. This fit, while strongly suggesting a physically reasonable cross-over from a predominance of phonon damping to a predominance of radiative damping, must therefore be considered phenomenological at this point. Furthermore, even if we assume that the temperature-independent damping is radiative, the existence of a threshold velocity below which the radiative damping does not occur needs explanation. Umklapp processes should permit radiative damping at all dislocation velocities, although work on a one-dimensional model indicates that as long as the potential is smooth, radiative damping is small for low velocities[24].

## 5. CONCLUSIONS

Based on molecular dynamics and classical EAM potentials, we have presented results on the velocity of edge and screw dislocations in Al and Ni, and edge dislocations in Al-Mg alloys as a function of applied stress and temperature.

In most cases, a regime exists where the velocity is linear in $\sigma/T$ (the applied shear stress divided by the temperature), as expected from theoretical studies of classical phonon damping. The behavior in this linear regime, for a given material and dislocation character, can therefore be captured by a single parameter $B/T$, which is shown in Table 3. Notably, Al screw dislocations are more strongly damped than Al edge dislocations, while damping in Ni is comparable between screw and edge dislocations. For Al, the calculated values of $B/T$ are consistent with the range of values found experimentally.

In the pure Al and Ni, the edge dislocation velocity remains a single, though nonlinear, function of $\sigma/T$ beyond the linear regime until the velocity saturates at a subsonic plateau. In contrast, for the screw dislocations there is a critical velocity above which the velocity has a temperature-independent component that can be fit to the radiation-damping form postulated by Eshelby. The behavior of the screw dislocations in the simulations at high subsonic velocities seems more in accord with what would be expected on theoretical grounds than that observed in the edge simulations, which show no evidence of a temperature independent, subsonic, radiative damping regime.

In the substitutional Al-Mg alloys, there is a regime of high temperature and intermediate stress where the velocity is also linear with $\sigma/T$ for edge dislocations. At lower stresses, the average velocity falls below that extrapolated from the linear regime towards the origin. This reduced velocity is, at least in part, caused by pinning of the dislocation by solute configurations. At lower temperatures (300K), the pinning effect is greater and the high-stress behavior quickly becomes non-linear. Thus at 300K, no regime with velocity linear in $\sigma/T$ is found.

At higher applied stresses and/or higher velocities, edge dislocations in the pure materials exhibit a subsonic plateau velocity and several transonic preferred velocities that have no direct relation to forbidden velocities in the materials. Screw dislocations can be driven to transonic velocities in Ni but not in Al. As the Ni screw dislocations become transonic, there is no evidence of the subsonic plateau that occurs for edge dislocations. At the highest applied stresses, the dislocations undergo a range of degradation processes. These high velocity results may be relevant to extreme loading conditions involving shock.

The above results could be used directly, or as guidance, in choosing dislocation mobility models for mesoscale discrete-dislocation simulations, such as those referenced in the introduction, or for coupled atomistic/discrete-dislocation models[43]. In situations where the velocities are not low, the present models demonstrate that edge and screw dislocations behave fundamentally differently. Such basic differences need to be incorporated into higher-scale constitutive models of dynamic behavior. Finally, the present computational methods and results provide a basis upon which to pursue more-detailed studies of the role of alloying elements in dislocation dynamics; such work will be reported in future publications.

## ACKNOWLEDGMENTS

This work was supported by General Motors through the GM/Brown Collaborative Laboratory on Computational Materials Science.

# REFERENCES

[1] J. Friedel, Dislocations, (Pergamon Press, New York, 1964).

[2] T.Y. Wu, J.L. Bassani and C. Laird, Proc. Roy. Soc. (London) **A435**, 1 (1991).

[3] N.A. Fleck and J.W. Hutchinson, Advances in Applied Mechanics **33**, 295 (1988).

[4] J. Lépinoux and L.P. Kubin, Scr. Metall., **21**, 833 (1987).

[5] A.N. Gulluoglu, D.J. Srolovitz, R. LeSar and P.S. Lomdahl, Scr. Metall., **23**, 1347 (1989).

[6] R. I. Amodeo and N.M. Ghoniem, Phys. Rev. B **41**, 6958 (1990); R. I. Amodeo and N.M. Ghoniem, Phys. Rev. B **41**, 6968 (1990).

[7] A.N. Gulluoglu and C.S. Hartley, Modelling Simul. Mater. Sci. Eng. **1**, 1 (1992); A.N. Gulluoglu and C.S. Hartley, Modelling Simul. Mater. Sci. Eng. **1**, 383 (1992).

[8] E. Van der Giessen and A. Needleman, Modelling Simul. Mater. Sci. Eng. **3**, 689 (1995).

[9] M.C. Fivel, T.J. Gosling and G.R. Canova, Modelling Simul. Mater. Sci. Eng. **4**, 581 (1996).

[10] M. Rhee, H.M. Zbib, J.P. Hirth, H. Huang and T. de la Rubia, Modelling Simul. Mater. Sci. Eng. **6**, 467 (1998).

[11] J.P. Hirth and J. Lothe, Theory of Dislocations, 2nd Ed. (John Wiley & Sons, New York, 1982).

[12] V.I. Alshits, in *Elastic Strain Fields and Dislocation Mobility*, edited by V.L. Indenbom and J. Lothe (North-Holland, Amsterdam, 1992).

[13] G. Leibfried, Z. Phys. **127**, 344 (1950).

[14] A.D. Brailsford, J. Appl. Phys. **41**, 4439 (1970); A.D. Brailsford, J. Appl. Phys. **43**, 1380 (1972).

[15] A. Hikata, R.A. Johnson, and C. Elbaum, Phys. Rev. B **2**, 4856 (1970); A. Hikata, R.A. Johnson, and C. Elbaum, Phys. Rev. B **4**, 674 (1971).

[16] J.D. Eshelby, Proc. Roy. Soc. **A62**, 307 (1949).

[17] F.C. Frank, Proc. Phys. Soc. **A62**, 131 (1949).

[18] A.N. Stroh, J. Math. Phys. **41**, 77 (1962).

[19] Three types of special velocities will be considered, any one of which might be referred to as a critical velocity. We shall avoid this term in the present developments.

[20] J.D. Eshelby, Proc. Phys. Soc. B **69**, 1013 (1956).

[21] H. Gao, Y. Huang, P. Gumbsch, and A.J. Rosakis, J. Mech. Phys. Sol. **47**, 1941 (1999).

[22] W. Atkinson and N. Cabrera, Phys. Rev. A **763**, 138 (1966).

[23] V. Celli and N. Flytzanis, J. Appl. Phys. **41**, 4443 (1970).

[24] S. Ishioka, J. Phys. Soc. Japan **34**, 462 (1973).

[25] V.I. Alshits and V.L. Indenbom, Sov. Phys. Usp. **18**, 1 (1975).

[26] F.R.N. Nabarro, Theory of Crystal Dislocations (Clarendon Press, Oxford, 1967).

[27] H.B. Huntington, J.E. Dickey, and R. Thomson, Phys. Rev. **113**, 1696 (1959) and many more.

[28] V.B. Shenoy and R. Phillips, Phil. Mag. A **76**, 367 (1997); S. Rao, C. Hernandez, J.P. Simmons, T.A. Parthasarathy and C. Woodward, Phil. Mag. A **77**, 231 (1998); V.V. Bulatov, O. Richmond and M.V. Glazov, Acta Mater. **47**, 3507 (1999); D.L. Olmsted, K.Y. Hardikar, and R. Phillips, Modelling Simul. Mater. Sci. Eng. **9**, 215 (2001); K.W. Jacobson and J. Schiotz, Nature Materials **1**, 15 (2002).

[29] D. Rodney and R. Phillips, Phys. Rev. Lett. **82**, 1704 (1999).

[30] M.S. Daw, M.I. Baskes, C.L. Bisson, and W.G .Wolfer, in *Modeling Environmental Effects on Crack Growth Processes: Proceedings of a Symposium*, edited by R.H. Jones and W.W. Gerberich (Metallurgical Society, Warrendale, PA, 1986), pp. 99-124.

[31] L.P. Grotelueschen, *Computer Simulation of Dislocation Dynamics in a Lennard-Jones Crystal Model*, Ph.D. Thesis, Brown University Division of Engineering, Providence RI., 1993.

[32] See http://www.fisica.uniud.it/~ercolessi/md/md/node15.html for a description of the Lennard-Jones potential.

[33] P. Gumbsch and H. Gao, Science **283**, 985 (1999); P. Gumbsch and H. Gao, J. Comput.-Aided Mater. Design **6**, 137 (1999).

[34] M. W. Finnis and J. E. Sinclair, Phil. Mag. **A50** 45 (1984).

[35] P. Rosakis, Phys. Rev. Lett. **86**, 95 (2001).

[36] N. Bhate, Computational and Experimental Studies of Dislocation Dynamics, Ph.D. Thesis, Brown University Division of Engineering, Providence, RI., 2001; N. Bhate, R.J. Clifton, and R. Phillips, in Shock Compression of Condensed Matter – 2001: Proceedings of the Conference of the American Physical Society Topical Group on Shock Compression of Condensed Matter, edited by M.D. Furnish, N.N. Thadani, and Y. Hori (American Institute of Physics, Melville, NY, 2002), pp. 339-342.

[37] F. Ercolessi and J.B. Adams, Europhys. Lett. **26**, 583 (1994).

[38] X.-Y. Liu, J.B. Adams, F. Ercolessi, J.A. Moriarty, Modelling Simul. Mater. Sci. Eng. **4**, 293 (1996); X.-Y. Liu, P.P. Ohotnicky, J.B. Adams, C.L. Rohrer, and R.W. Hyland, Jr., Surf. Sci. **373**, 357 (1997).

[39] See **http://ceaspub.eas.asu.edu/cms/welcome.htm** for available EAM potentials and associated descriptions.

[40] E Nadgornyi, Progr. Materials Sci., **31**, 1 (1988).

[41] J. A. Gorman, D. S. Wood and T Vreeland, J. Appl. Phys. **40**, 833 (1969); J. Appl. Phys. **40**, 903 (1969).

[42] V. R. Parameswaran, N. Urabe and J. Weertman, J. Appl. Phys. **43**, 2982 (1972); V. R. Parameswaran and J. Weertman, Met. Trans. **2**, 1233 (1971).

[43] L.E. Shilkrot, R.E. Miller, and W.A. Curtin, Phys. Rev. Lett., **89**, 025501 (2002).

# APPENDIX
# SIMULATION DETAILS

The slip plane of the dislocation is located at the mid-point of the simulation cell in the y-direction, with the slip plane lying along the negative x-direction. The displacements at the boundaries are established from a notional larger simulation cell containing an additional dislocation of opposite sign below the actual dislocation, displaced in the y-direction to create a dislocation dipole, with periodic boundary conditions in the y-direction.

The displacements at the cell boundary may not be initially obvious, and hence we describe these in some detail here. Near the positive x-boundary, the displacements are linearly interpolated from the elasticity displacements in an infinite medium to zero displacement at the boundary, independent of y. At the negative x-boundary, the displacements are interpolated to **b**/2 above the slip plane, and $-$**b**/2 below the slip plane where **b** is the Burgers vector. This leaves a discrepancy between the negative and positive x-boundaries in violation of the periodic boundary conditions. For the edge dislocation, where **b** is in the x-direction, this represents the excess and/or missing atoms that result from insertion of an edge dislocation with its cut along the slip plane. With the Burgers vector taken in the x-direction, the displacement of **b**/2 above the slip plane at the negative x-boundary requires the insertion of a half-plane of atoms above the slip plane, and similarly, a half plane of atoms below the slip plane is deleted (this could be considered, equivalently, as deleting two half planes of atoms below the slip plane). The planes have an AB stacking order, and by choosing an even number of planes of atoms above the slip plane (and hence an even number below the slip plane), the atoms are correctly aligned across the x-boundary.

For the screw dislocation with Burgers vector in the z-direction, no atoms need to be added or removed. The stacking order of the planes in the x-direction is ABCDEF, and if the number of planes is divisible by 6, then the displacements of $\pm b/2$ would cause the planes to be misaligned at the x-boundary. The number of planes is therefore taken as $6N_x + 3$ for some $N_x$, and since the displacement of $\pm b/2$ turns a C-plane into an F-plane; for example, the crystal is now correctly matched at the boundary.

For all of the simulations, except the pure aluminum edge dislocation simulations, the size of the simulation cell was determined by conducting a simulation at constant pressure. For this purpose, a cell that was twice the height (i.e. the y-direction) of the desired simulation cell was built, with full periodic boundary conditions and a dislocation dipole. The dislocation of opposite sign was offset in the y-direction from the regular dislocation. In this case, there was no need for any constraints on the atomic positions. After relaxing the dislocation, a simulation was conducted at constant temperature and pressure, with the lengths $L_x$, $L_y$, and $L_z$ allowed to fluctuate. The average lengths were then used to determine the shape of the desired cell with only one dislocation. If the material were purely isotropic, this would eliminate all of the overall stress in the cell. Since the shape of the cell was kept rectangular, a residual shear stress remained due to elastic anisotropy. This is a y-z stress in the case of an edge dislocation, and it is an x-y stress in the case of the screw dislocation. In both cases, the residual shear stress does not exert a Peach-Koehler force on the dislocation.

Table 1. Elastic data at 0K for the EAM potentials used.

| Property | Al[a] | Ni[b] |
|---|---|---|
| Lattice constant (Å) | 4.032 | 3.52 |
| Elastic moduli (GPa) | | |
| $C_{11}$ | 118.0 | 245 |
| $C_{12}$ | 62.20 | 148 |
| $C_{44}$ | 36.71 | 134 |
| Anisotropy ratio[11] $\left(\dfrac{2C_{44}}{C_{11}-C_{12}}\right)$ | 1.3 | 2.8 |
| Wave speeds ($10^3$ m/s) | | |
| Direction of motion of edge dislocation $[1,\bar{1},0]$ | | |
| Longitudinal | 6.81 | 6.08 |
| Transverse | 3.66 | 3.87 |
| | 3.18 | 2.33 |
| Direction of motion of screw dislocation $[1,1,\bar{2}]$ | | |
| Quasi-Longitudinal | 6.82 | 6.14 |
| Quasi-Transverse | 3.51 | 3.43 |
| | 3.35 | 2.81 |
| Forbidden speeds ($10^3$ m/s) | | |
| Edge dislocation: identical to the wave speeds | | |
| Screw dislocation | | |
| | 6.33 | 5.68 |
| | 3.48 | 3.07 |
| | 3.33 | 2.77 |

[a] Lattice constant and elastic moduli are as computed in our simulations, and are in good agreement with Ref. 37.

[b] Lattice constant and elastic moduli are as computed in our simulations. The potential is from Ref. 30.

Table 2. Simulation cell dimensions and total # atoms.

| System | Typical Simulation Cell Dimensions for 300K (Å) | | | #Atoms |
|---|---|---|---|---|
| | $L_x$ | $L_y$ | $L_z$ | |
| Al edge | 113.2 | 91.2 | 49.6 | 31,600 |
| Al screw | 116.6 | 81.9 | 28.6 | 16,920 |
| Ni edge | 98.8 | 79.5 | 43.2 | 31,600 |
| Ni screw | 101.7 | 79.6 | 25.0 | 18,800 |
| Al2.5at%Mg | 113.6 | 91.5 | 99.5 | 63,200 |
| Al5.0at%Mg | 114.0 | 91.8 | 99.8 | 63,200 |

Table 3. Damping fit as linear in temperature

| B/T ($10^{-8}$ Pa s / K)[a] | Data up to 0.3 MPa / K | | Data up to 0.15 MPa / K | |
|---|---|---|---|---|
| | Al | Ni | Al | Ni |
| Edge dislocation | 3.9 | 5.0 | 3.7 | 5.3 |
| Screw Dislocation | 7.5 | 6.4 | 6.0 | 5.5 |

[a] T/B was fit by least squares as described in the text.

Table 4. Damping fit as linear in stress for alloys

| B/T ($10^{-8}$ Pa s / K)[b] | Al2.5at%Mg | Al5.0%Mg |
|---|---|---|
| Edge dislocation | 5.0 | 5.5 |

[b] As given by the freehand fits to the 500K data shown in Fig. 4. These are fits to intermediate stress data where pinning does not occur. Note that the 700K data agree with the 500K freehand fit.

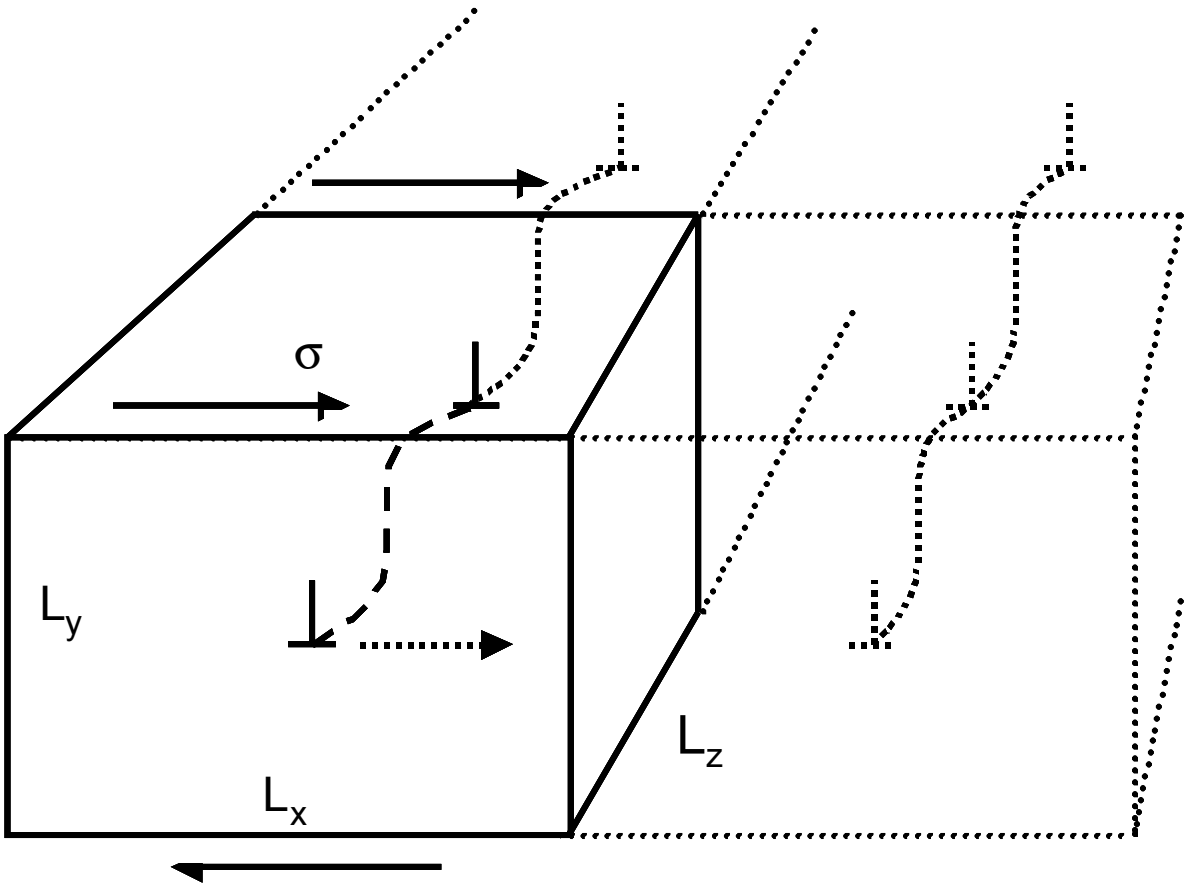

Figure 1. Simulation cell of sides $L_x$, $L_y$, $L_z$ with single dislocation subject to an applied shear stress $\sigma$.

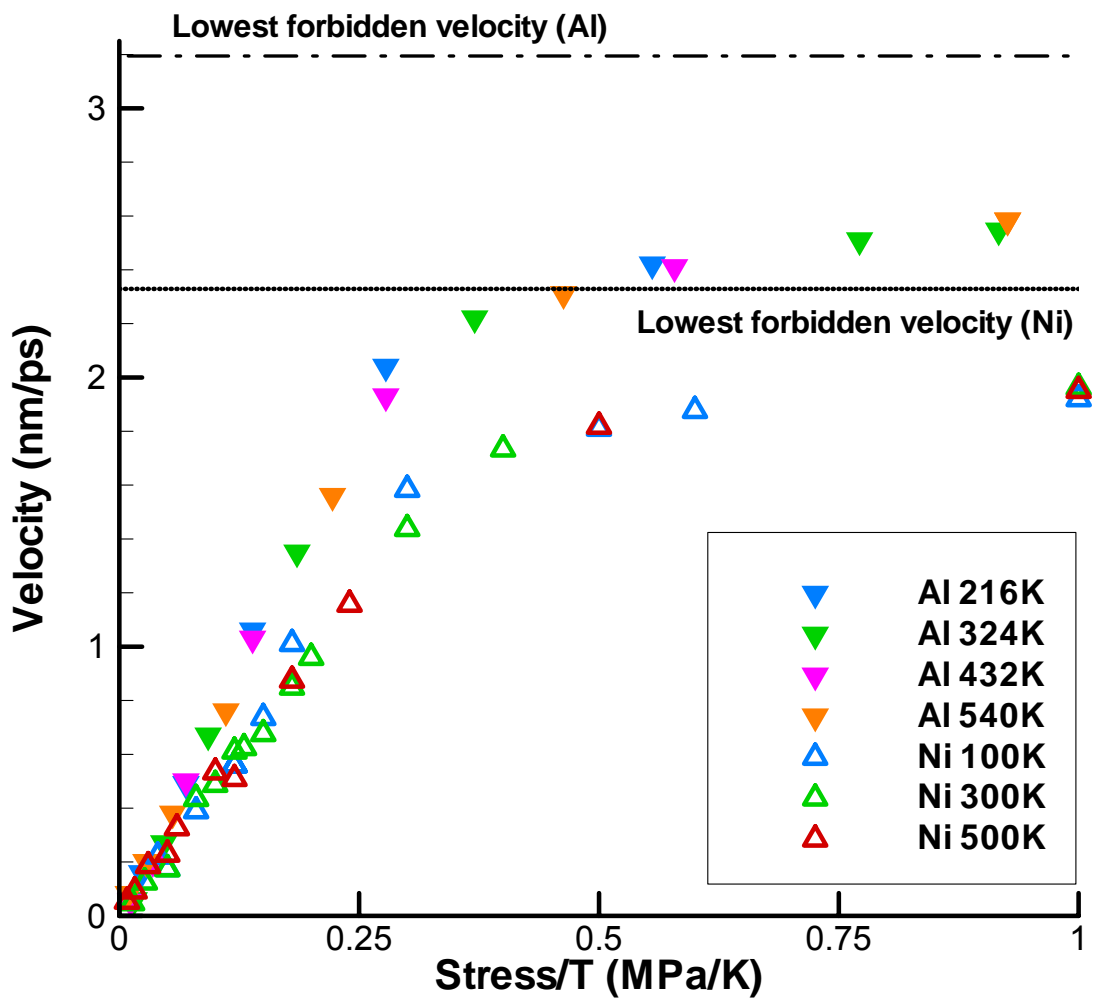

Figure 2. Velocity as a function of the applied shear stress divided by the temperature ($\sigma/T$) for edge dislocations in Al and Ni. The velocity is nearly linear for $\sigma/T < 0.3$ MPa/K, and remains a function of $\sigma/T$ up to $\sigma/T = 1.0$ MPa/K, but saturates at a value below the lowest forbidden velocity.

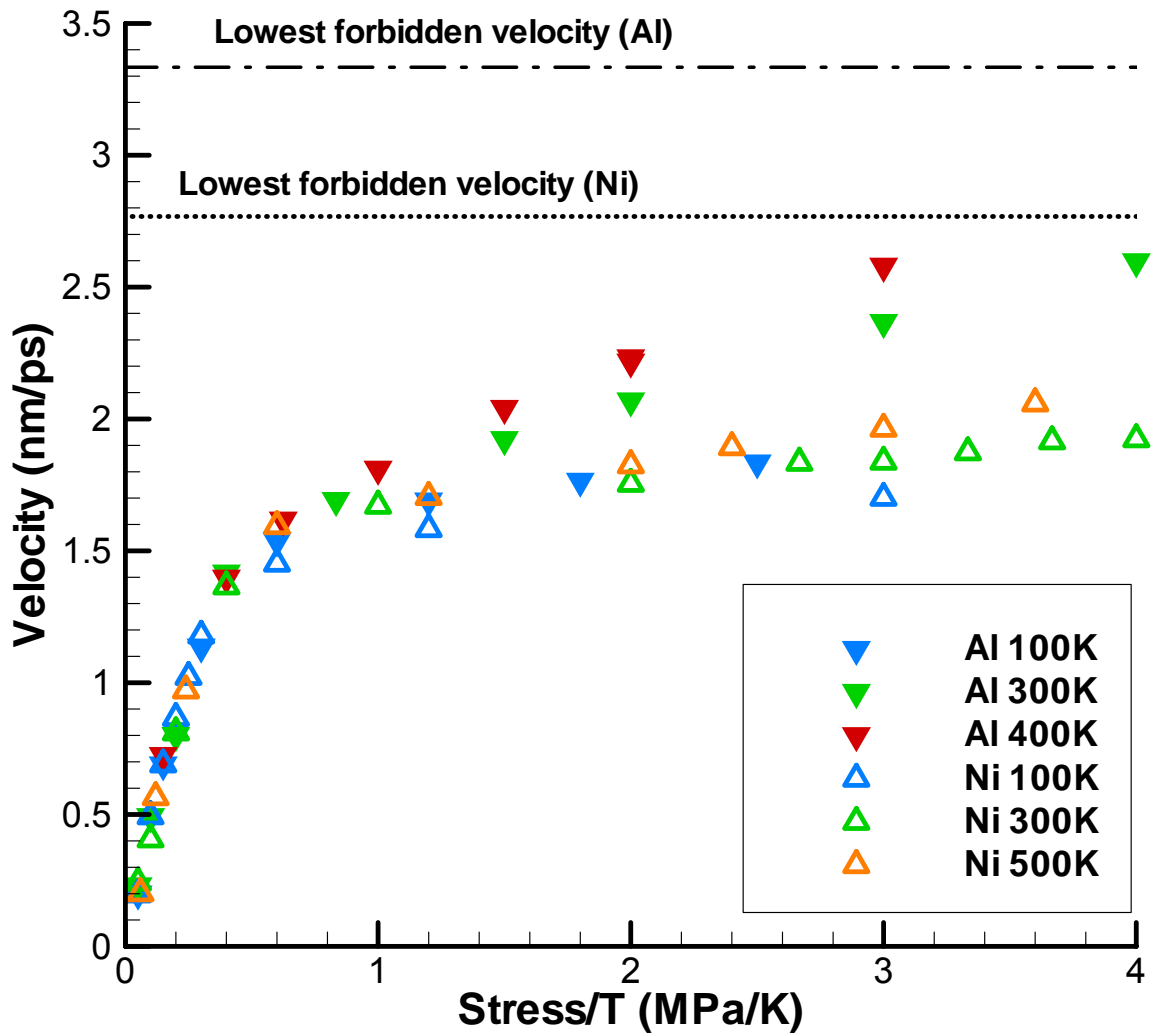

Figure 3. Velocity as a function of the applied shear stress divided by the temperature (σ/T) for screw dislocations in Al and Ni. For σ/T < 0.4 MPa/K, the velocity is roughly linear and depends only on σ/T. For σ/T > 0.6 MPa/K, the velocity no longer depends solely on σ/T.

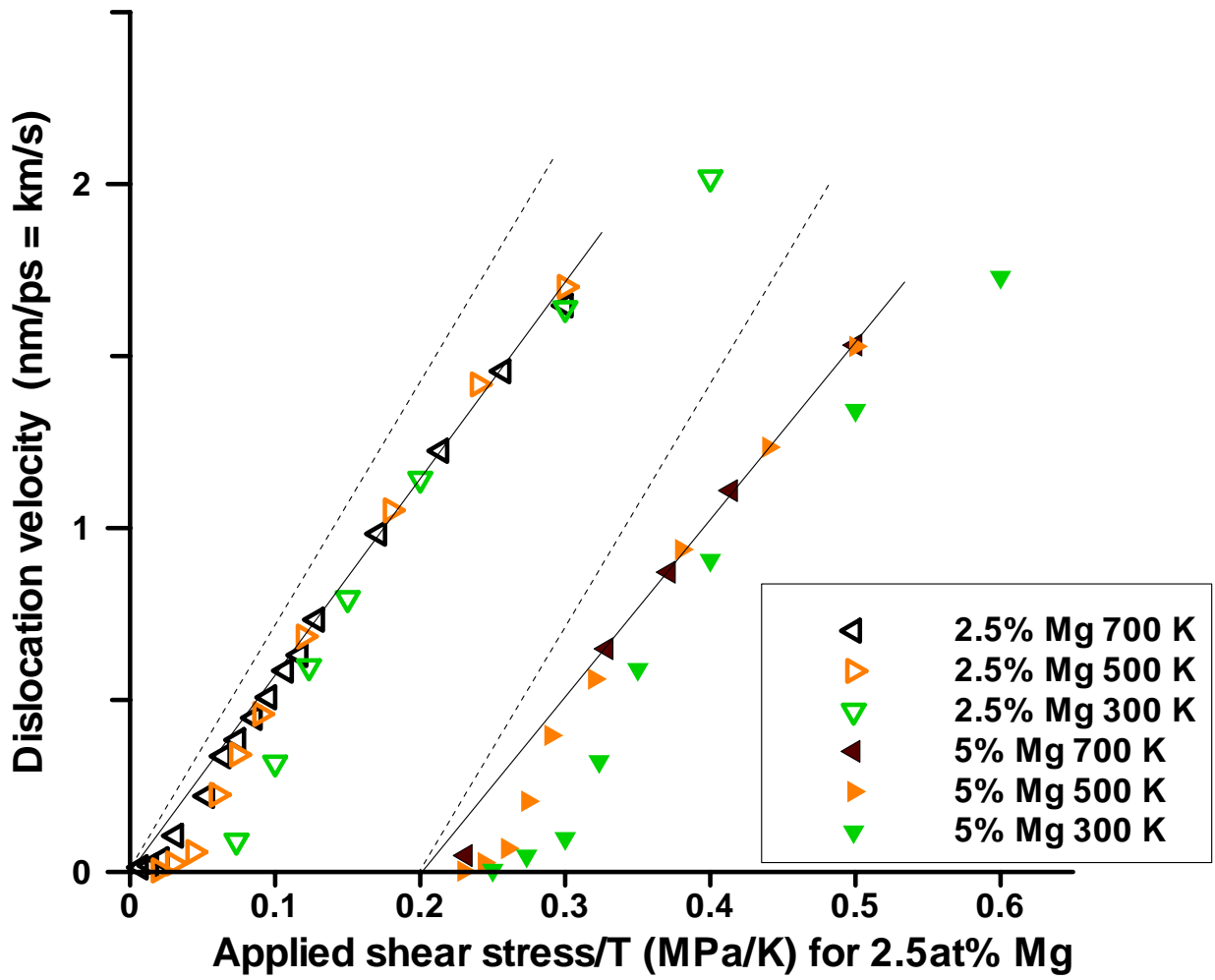

Figure 4. Velocity as a function of the applied shear stress divided by the temperature (σ/T) for edge dislocations in Al-Mg random substitutional alloys (2.5at%Mg and 5.0at%Mg are shown). The data for Al5.0at%Mg is offset by 0.2 MPa/K for clarity. Solid lines: linear fits to 500K data. Dashed lines: fit for pure Al edge data, for reference.

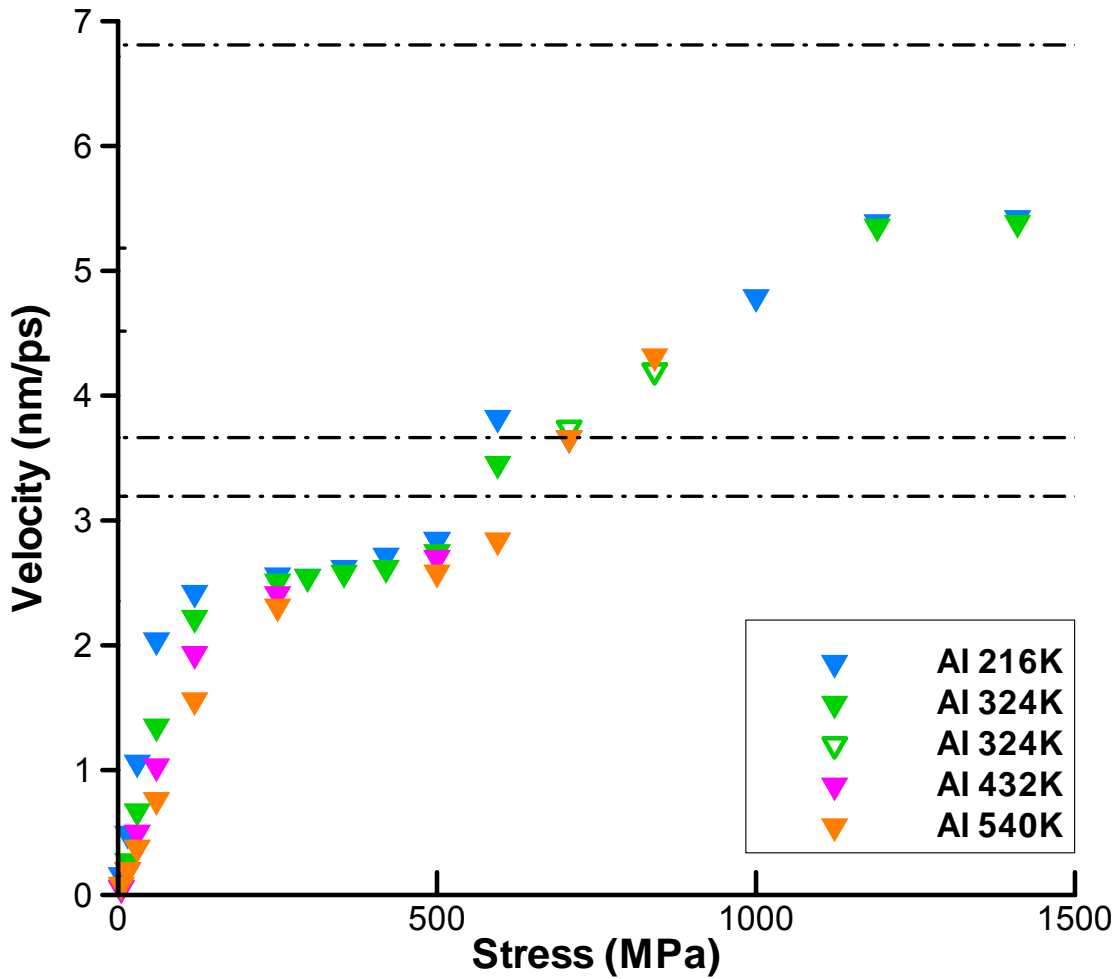

Figure 5. Velocity as a function of the applied shear stress for an edge dislocation in aluminum. The dashed lines are the forbidden velocities from continuum linear elasticity, which for the edge dislocation are equal to the wave speeds. For simulations with σ/T at least about 0.9, and σ no more than about 400 MPa there is a plateau region where the velocity saturates at about 2.6 nm/ps, independent of temperature. There is some evidence of a jump across the velocity region of at least the lower of the two similar shear wave speeds. The open symbols for 324K for 707 MPa and 841 MPa are described in the text.

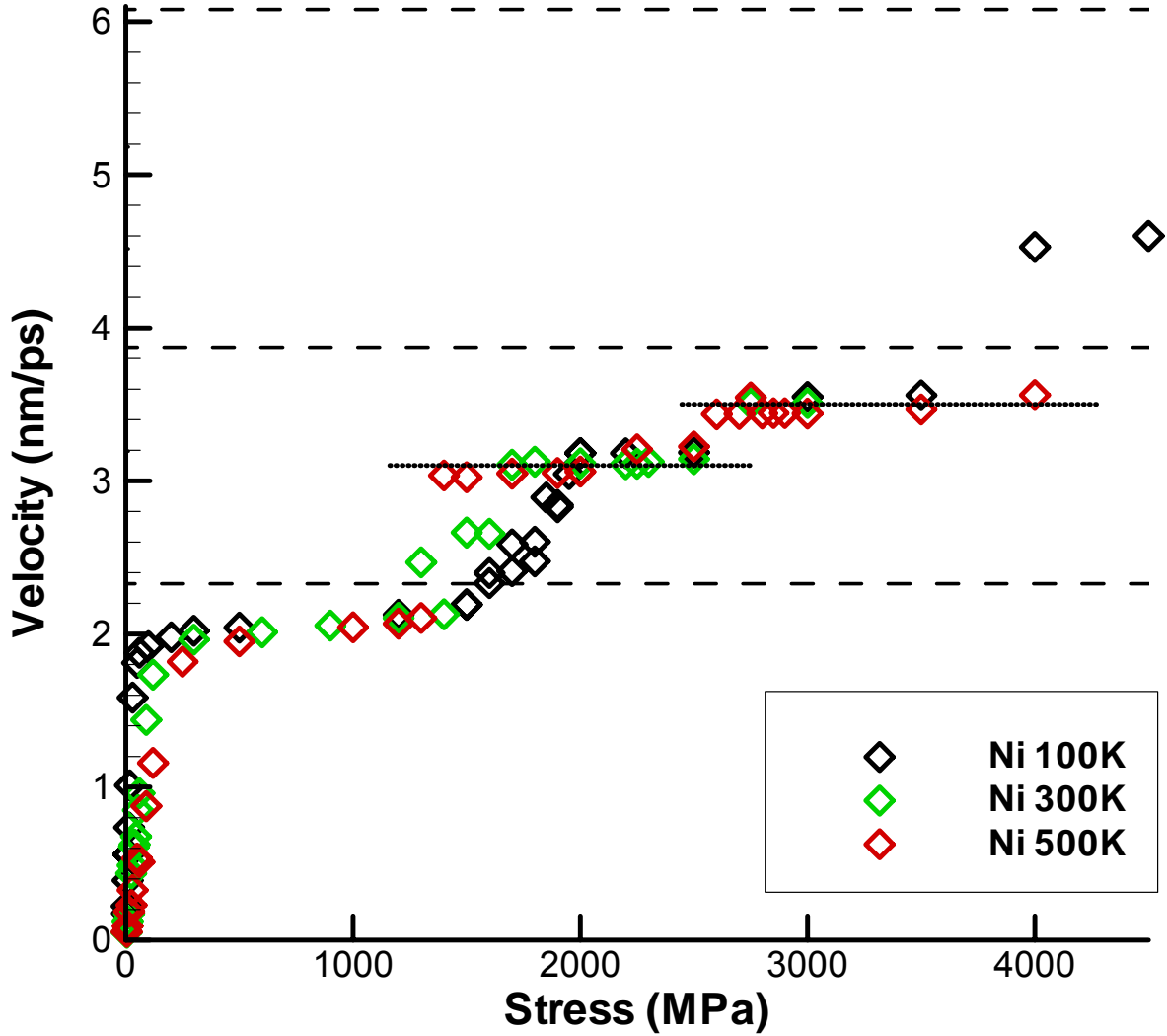

Figure 6. Velocity as a function of the applied shear stress for an edge dislocation in nickel. The dashed lines are the forbidden velocities from continuum linear elasticity, which for the edge dislocation are equal to the wave speeds. For simulations with σ/T at least about 1.0, and σ no more than about 1 GPa there is a plateau region where the velocity saturates at about 2.0-2.1 nm/ps, independent of temperature. There is then some evidence of a jump across the lower shear wave speed, with preferred velocities around 3.0-3.2 nm/ps (around 1.5 GPa to 2.5 GPa) and 3.4-3.6 nm/ps (around 2.5 GPa to 4.0 GPa). Dotted horizontal lines at 3.1 nm/ps and 3.5 nm/ps are intended only to guide the eye. Finally, the 100 K simulations at 4.0 GPa and 4.5 GPa have jumped above the higher shear wave speed.

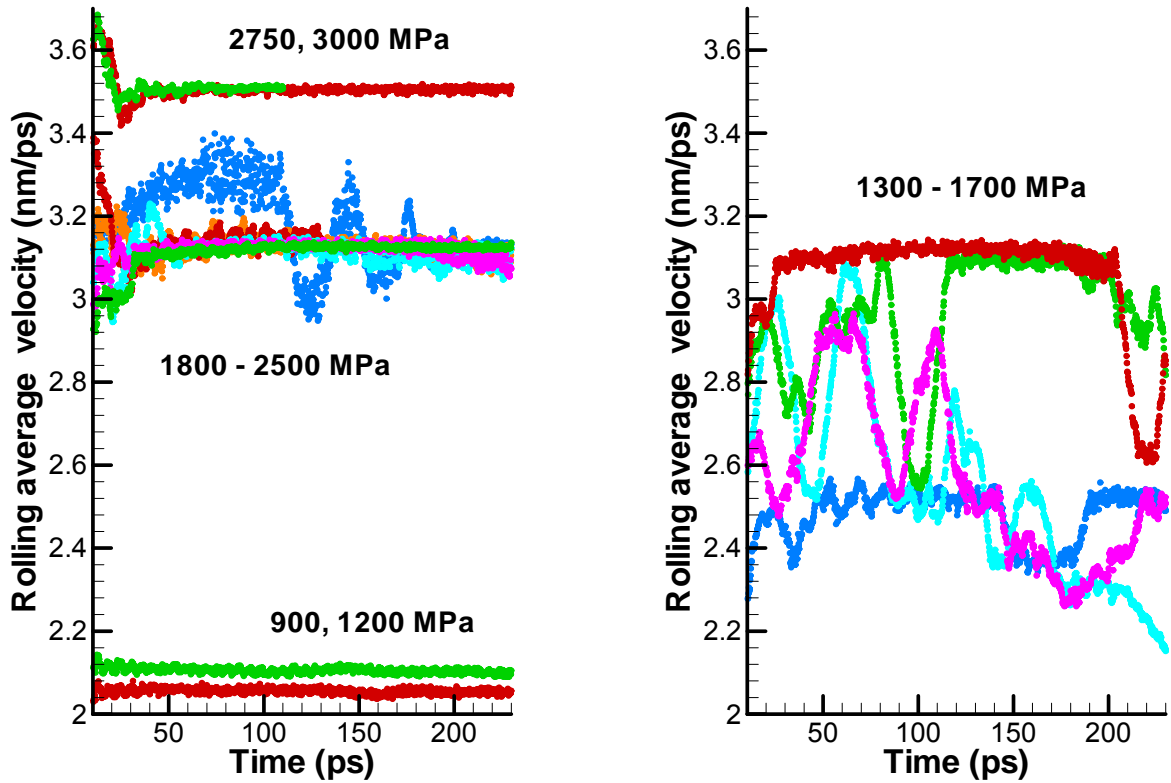

Figure 7. Ni edge dislocation velocity as a function of time. Shown is the average velocity over a rolling 20 ps period. At 900 and 1200 MPa the average velocity is subsonic and relatively steady. From 1300 to 1700 MPa (shown separately on the right for clarity) the velocity is much less steady. For higher stresses there are preferred velocities at 3.1 and 3.5 nm/ps. The shift from the behavior at 900 and 1200 MPa and that from 1300-1700 MPa is consistent with the continuum elasticity prediction that there is a forbidden velocity at 2.33 nm/ps.

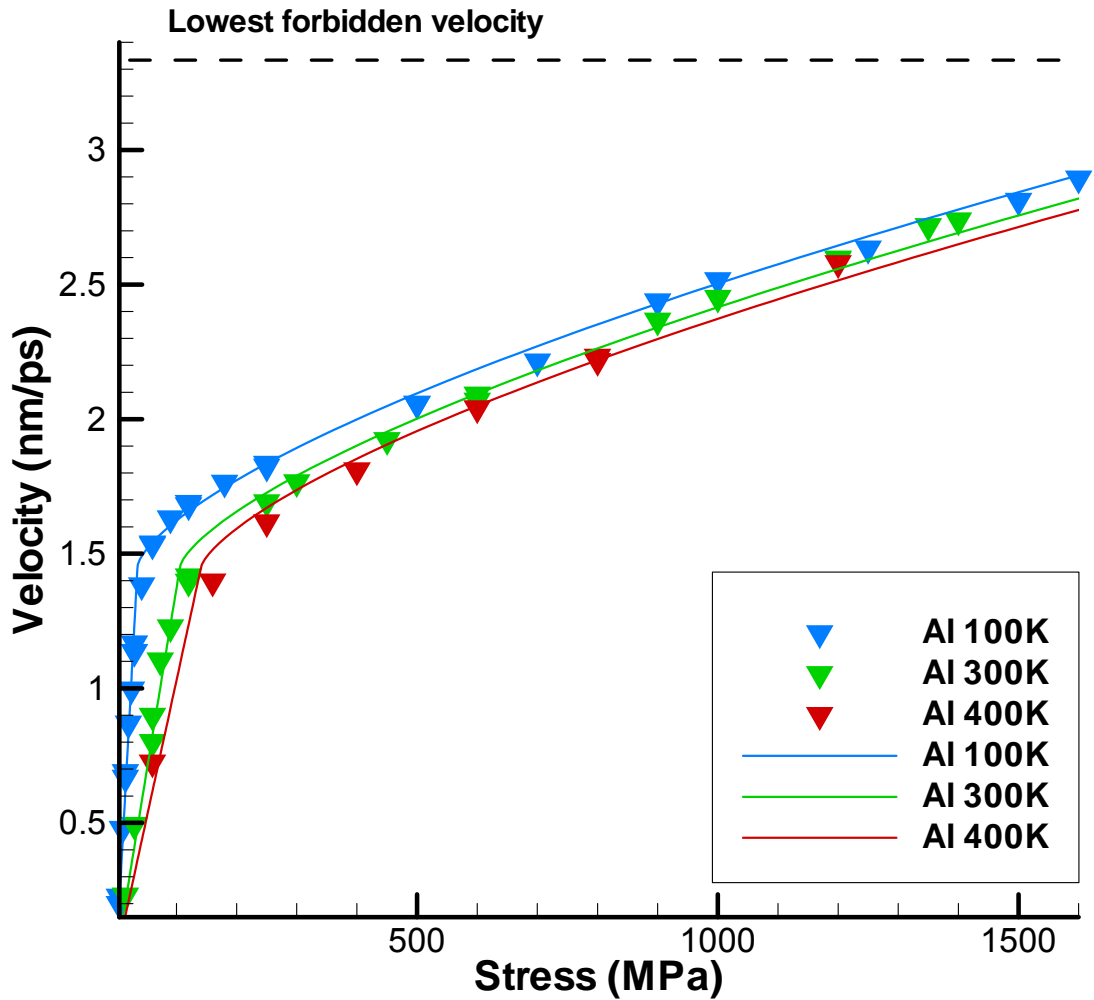

Figure 8. Aluminum screw dislocation velocity as a function of the applied shear stress. The lines are a fit of the data to Eq. (3).

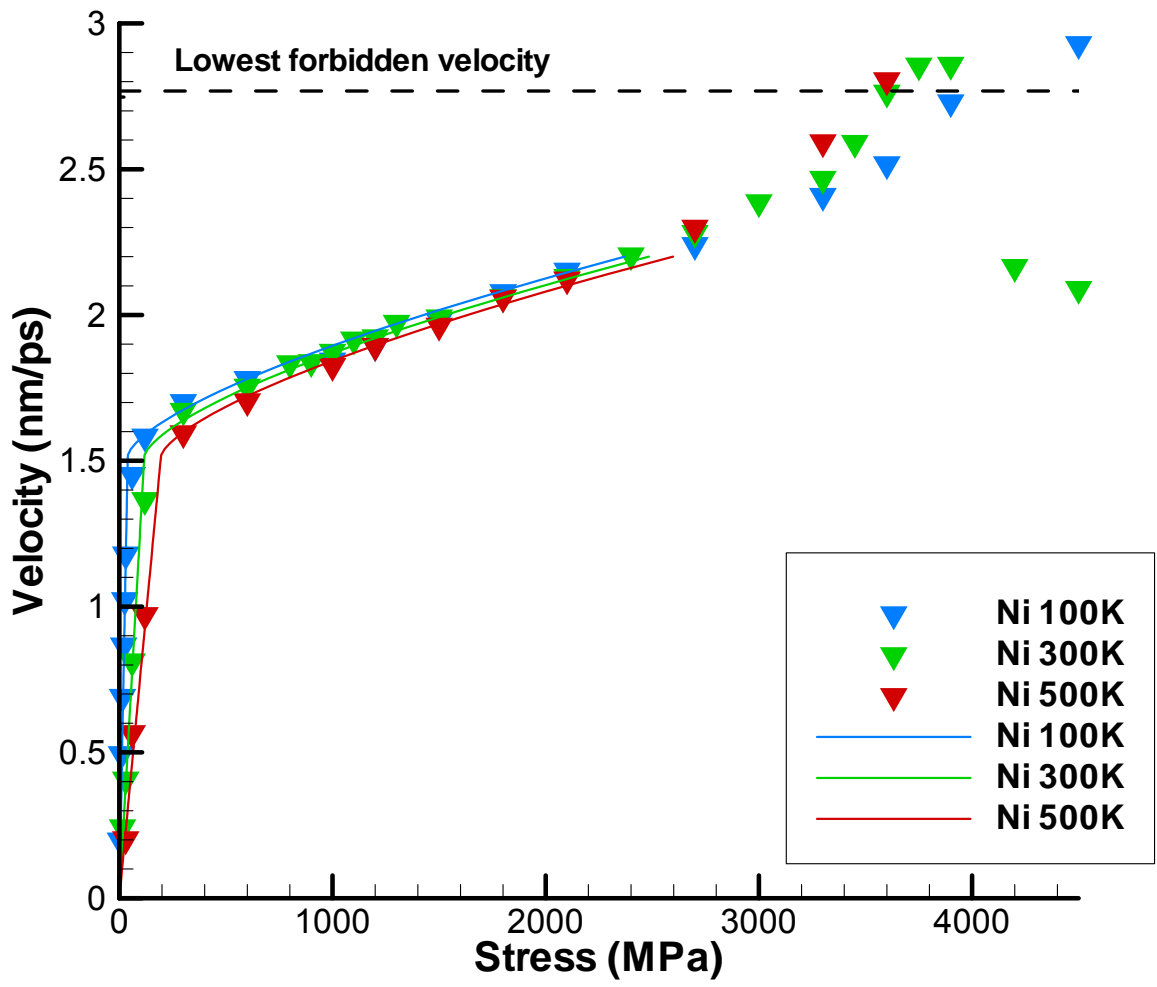

Figure 9. Nickel screw dislocation velocity as a function of the applied shear stress. The lines are a fit of the data to Eq. (3).